\documentclass[twocolumn]{webofc}

\usepackage[varg]{txfonts}   
\usepackage{hyperref}
\usepackage{url}

\newcommand{\dd}{\mathrm{d}}

\hypersetup{colorlinks=true,citecolor=blue,urlcolor=blue,linkcolor=blue}
\begin{document}
\title{Proton-neutron correlations in baryon-number fluctuations near the liquid-gas transition}
%
%

\author{\firstname{Michał} \lastname{Marczenko}\inst{1,2}\fnsep\thanks{\email{michal.marczenko@uwr.edu.pl}}
}

\institute{%
Institute of Theoretical Physics,\\ University of Wroc\l{}aw, 
plac Maksa Borna 9, 50-204 Wroc\l{}aw, Poland
\and%
Incubator of Scientific Excellence - Centre for Simulations of Superdense Fluids,\\ University of Wroc\l{}aw, 
plac Maksa Borna 9, 50-204 Wroc\l{}aw, Poland
}

\abstract{
We study net-baryon number density fluctuations in isospin-symmetric matter near the nuclear liquid-gas phase transition using the parity doublet model. We analyze second-order susceptibilities of net-proton and net-neutron numbers and their correlations. We show that proton-neutron correlations are nontrivial and lead to qualitative differences between net-proton and net-baryon fluctuations. We further investigate factorial cumulants and demonstrate that the differences between baryon- and proton-number factorial cumulants are governed by proton-neutron correlations. Our results highlight the importance of interaction-driven correlations for interpreting fluctuation measurements near the liquid-gas critical endpoint.
}

\maketitle

\section{Introduction}
\label{sec:intro}

One of the central goals of high-energy physics is to map the phase diagram of quantum chromodynamics (QCD)~\cite{Stephanov:2024xkn}. Lattice QCD studies have established that at finite temperature and vanishing chemical potential the transition from hadronic matter to quark-gluon plasma is a smooth crossover~\cite{HotQCD:2014kol, Borsanyi:2018grb}. At finite chemical potential, the phase structure is expected to be richer. Effective models, Dyson-Schwinger approaches, and functional renormalization group studies indicate the possible existence of a first-order phase transition and a critical point in the QCD phase diagram~\cite{Qin:2010nq}.

Locating a possible QCD critical point is a major goal of large-scale heavy-ion collision experiments. It has been proposed that the QCD phase structure can be probed through fluctuations and correlations of conserved charges, which are sensitive to critical behavior near the phase boundary~\cite{Friman:2011pf}. In particular, nonmonotonic behavior of net-baryon cumulants has been suggested as a signature of the QCD critical point and remnants of chiral criticality~\cite{Stephanov:2011pb}. Results from the Beam Energy Scan program at RHIC show indications of nonmonotonic behavior in net-proton cumulants~\cite{Luo:2017faz}, but no definitive conclusion has yet been reached, highlighting the need for higher-statistics measurements at low collision energies.

It has been argued that at large baryon chemical potentials the chemical freeze-out line may approach the critical point of the nuclear liquid-gas phase transition~\cite{Floerchinger:2012xd}. Low-energy heavy-ion experiments, such as those by the HADES Collaboration at GSI, aim to probe this region through fluctuation measurements. The properties of nuclear matter and the nuclear liquid-gas transition have been extensively studied both experimentally and theoretically~\cite{Hempel:2013tfa, Koch:2023oez, Marczenko:2023ohi, Marczenko:2021icv, Marczenko:2024jzn}.

Due to experimental limitations, only charged particles can be measured in heavy-ion collisions, and net-proton fluctuations are therefore commonly used as a proxy for net-baryon fluctuations. This identification is valid in an ideal gas framework but can be violated by interaction-induced correlations. At low collision energies, where hadronic interactions dominate, net-proton fluctuations may cease to reliably reflect net-baryon fluctuations. While qualitative differences between net-baryon and baryonic-sector fluctuations have been studied near the liquid-gas and chiral transitions, a systematic understanding of the relation between net-proton and net-baryon fluctuations near the liquid-gas transition remains incomplete.

We use the parity doublet model to study net-proton and net-neutron fluctuations and their correlations near the nuclear liquid-gas phase transition. We focus on second-order susceptibilities and factorial cumulants to elucidate the role of proton-neutron correlations in baryon number fluctuation observables.

\section{Fluctuations near the nuclear liquid-gas phase transition}
\label{sec:fluct}

\begin{figure}[t]
    \centering
    \includegraphics[width=\columnwidth]{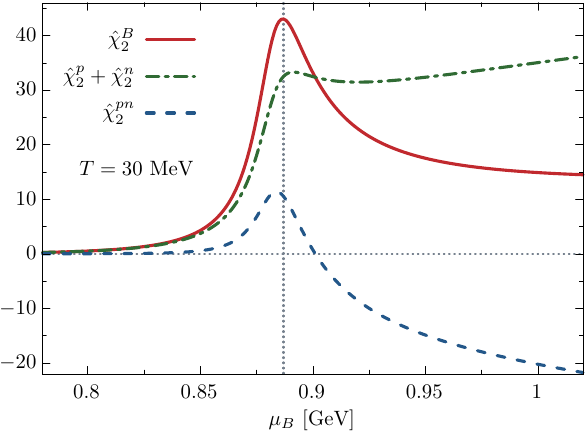}
    \caption{Net-baryon number susceptibility at $T=30~\rm MeV$ in the vicinity of the liquid-gas phase transition. The grey, dotted vertical line marks the crossover transition obtained from the maximum of $\hat \chi_2^B$.}
    \label{fig:x2_t30}
\end{figure}

In this work, we model the liquid-gas phase transition with the hadronic parity doublet model in the mean-field approximation (see~\cite{Marczenko:2024nge} for details). We consider a two-flavor isospin-symmetric system and pay attention to the fluctuations near the critical point of the liquid-gas phase transition. Thus, we restrict our considerations to low temperatures and small densities, where the system is predominantly composed of positive-parity nucleons. Therefore, in this region, the thermodynamic potential can be reliably restricted to
\begin{equation}\label{eq:thermo_lg}
    \Omega = \Omega_p + \Omega_n + V_\sigma + V_\omega \textrm,
\end{equation}
where $\Omega_{p}(\Omega_n)$ is the potential of protons (neutrons) and $V_\omega$ contains the mesonic mean-field potentials. The net-baryon number density is
\begin{equation}
    n_B = -\frac{\dd \Omega}{\dd \mu_B}\Bigg|_T = n_p + n_n \textrm, 
\end{equation}
where $n_{p}~(n_n)$ are the net-proton (net-neutron) number densities. Due to isospin symmetry $n_p = n_n = 1/2~n_B$.

The generalized susceptibilities of the net-baryon number, $\chi_k^B$, are defined as derivatives with respect to the baryon chemical potential,
\begin{equation}
    \hat \chi_k^B = -\frac{\dd^k \hat\Omega}{\dd \hat\mu_B^k}\Bigg|_T \textrm,
\end{equation}
where the hat symbol indicates that a quantity is temperature-normalized. The second-order susceptibility of the net-baryon number can be written as follows
\begin{equation}\label{eq:x2_pn}
    \hat\chi_{2}^B = \hat\chi_2^{p}  + \hat\chi_2^{n} + \hat\chi_2^{pn}\textrm,
\end{equation}
where $\hat\chi_2^{p}$ and $\hat\chi_2^{n}$ are the susceptibilities of the net-proton and net-neutron numbers (see~\cite{Marczenko:2024nge} for a detailed derivation),
\begin{equation}\label{eq:susc_2}
    \hat \chi_2^{p/n} = -\frac{\dd^2 \hat \Omega}{\dd \hat\mu_{p/n}^2}\Bigg|_{T}\textrm,
\end{equation}
respectively. We note that $\hat\chi_2^{p} = \hat \chi_2^{n}$, due to isospin symmetry assumed in this work. The last term, $\hat\chi_2^{pn}$, is the correlation between the net-proton and net-neutron numbers,
\begin{equation}\label{eq:corr_2}
    \hat \chi_2^{pn} = -2\frac{\dd^2 \hat \Omega}{\dd \hat\mu_p \,\dd \hat\mu_n}\Bigg|_{T}\textrm.
\end{equation}

\begin{figure}[t]
    \centering
    \includegraphics[width=\columnwidth]{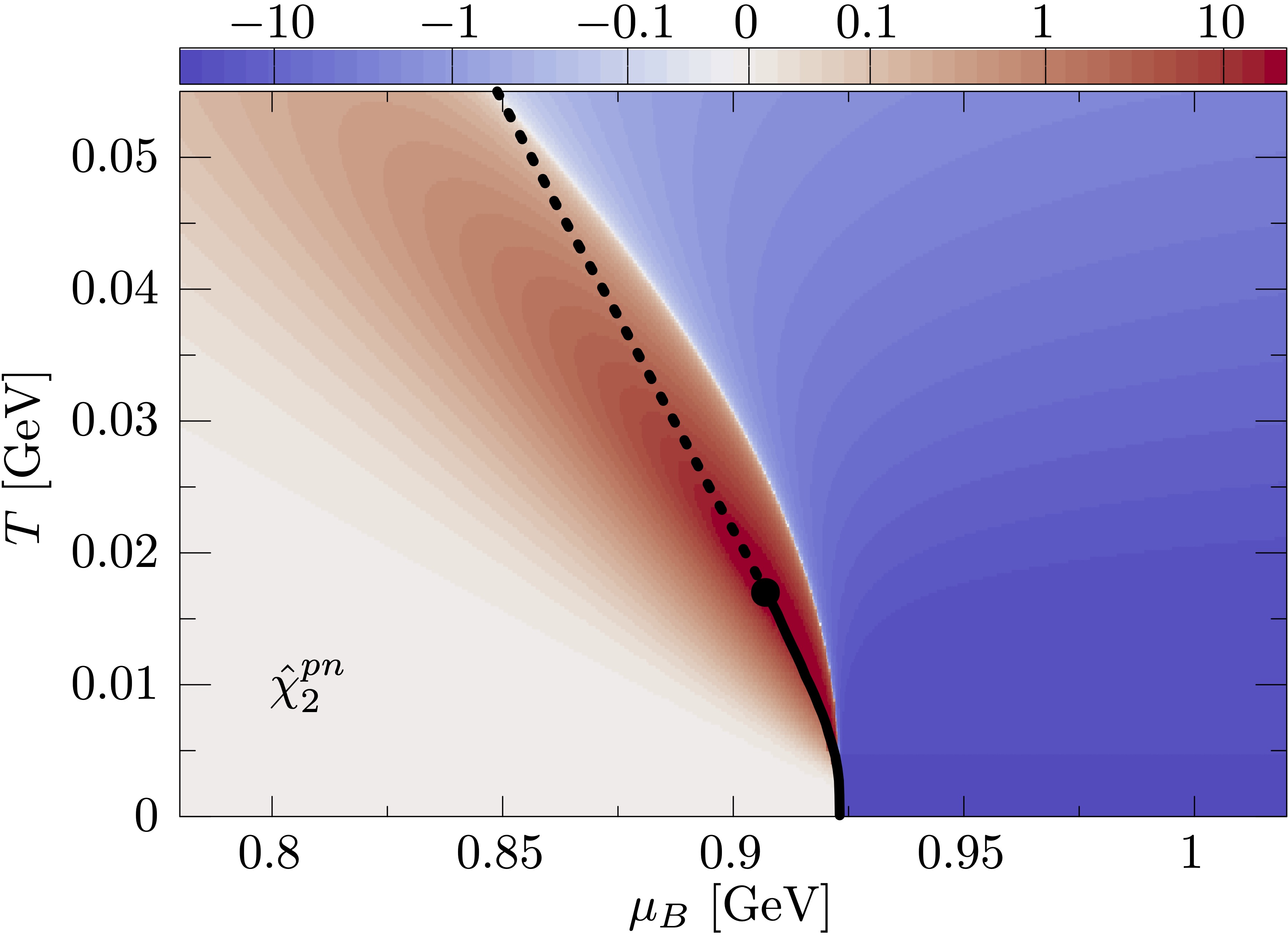}
    \caption{The proton-neutron correlator $\hat\chi_2^{pn}$. The black, solid line marks the first-order liquid-gas phase transition ending with a circle that indicates the critical point. The black, dotted line indicates the crossover transition obtained from the maximum of $\hat\chi_2^B$.}
    \label{fig:heatmap_corr}
\end{figure} 

Figure~\ref{fig:x2_t30} shows the net-baryon susceptibility $\hat\chi_2^B$ at $T=30~\mathrm{MeV}$ near the liquid-gas transition. It exhibits a clear maximum across the crossover and decreases at larger baryon chemical potential. The proton susceptibility $\hat\chi_2^{p}$ follows a similar trend at low chemical potential but begins to deviate near the transition and shows a qualitatively different behavior afterward. This deviation signals a nonvanishing proton-neutron correlator $\hat\chi_2^{pn}$~[cf.~Eq.~\eqref{eq:x2_pn}]. At low chemical potential, $\hat\chi_2^{pn}\simeq0$ and $\hat\chi_2^B\simeq2\hat\chi_2^{p}$. Close to the crossover, however, the correlator develops a pronounced structure, increasing toward a peak near the transition and becoming negative at larger chemical potential. This leads to an enhancement of $\hat\chi_2^{p}$ relative to $\hat\chi_2^B$ beyond the transition. Since net-proton fluctuations are experimentally used as proxies for net-baryon fluctuations, a vanishing $\hat\chi_2^{pn}$ is implicitly assumed. Our results show that this is not the case near the critical region, consistent with nonzero off-diagonal susceptibilities observed in lattice QCD and effective model studies~\cite{Gavai:2005yk}. 

Figure~\ref{fig:heatmap_corr} displays the proton-neutron correlator $\hat\chi_2^{pn}$ in the $\mu_B$-$T$ plane. For all temperatures shown, the correlator vanishes as $\mu_B \rightarrow 0$, reflecting free-gas behavior. In the vicinity of the critical point, $\hat\chi_2^{pn}$ becomes sizable, develops a maximum, and turns negative at higher baryon chemical potential. This nontrivial structure shows that the common identification of net-proton fluctuations with half of the net-baryon fluctuations breaks down near the critical region. The result is obtained for isospin-symmetric matter with complete isospin randomization, $n_B = 2n_{p} = 2n_n$.

\begin{figure}[t]
    \centering
    \includegraphics[width=1\linewidth]{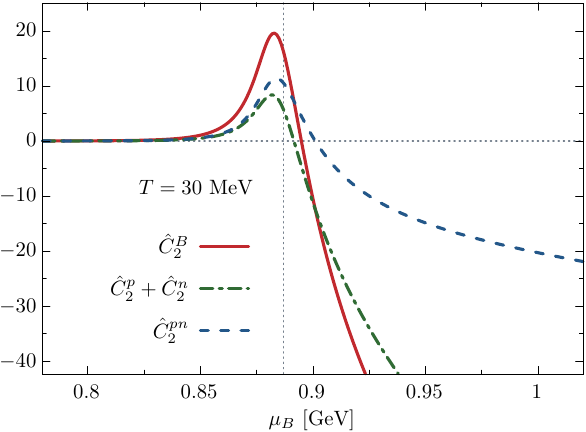}
    \caption{Second-order factorial cumulants, $\hat C^\alpha_2$, at $T=30~\rm MeV$ in the vicinity of the liquid-gas phase transition. The grey, dotted vertical line marks the crossover transition obtained from the maximum of $\hat \chi_2^B$.}
    \label{fig:C2}
\end{figure}

Nonmonotonic behavior is also expected for ratios of higher-order net-baryon fluctuations, which are commonly studied in experiments since they reduce sensitivity to volume effects. Ratios of the net-baryon number exhibit characteristic structures near the liquid-gas phase boundary-consistent with expectations from universality arguments~\cite{Stephanov:2011pb, Stephanov:2008qz}. However, the corresponding ratios constructed from net-proton fluctuations show a qualitatively different behavior. In particular, the region where critical structures appear is significantly reduced when net-proton observables are considered instead of net-baryon ones. This is why the proton-number fluctuation ratios may substantially underestimate the extent of the critical region, highlighting the importance of baryonic correlations for interpreting experimental fluctuation measurements~\cite{Marczenko:2024nge}.

\section{Factorial Cumulants}

Baryon number cumulants can, in principle, be reconstructed from experimentally measured proton cumulants under the assumption of complete isospin randomization in the final state~\cite{Kitazawa:2012at}. In this framework, factorial cumulants are particularly useful, as they remove self-correlations and lead to a simple proportionality between proton and baryon factorial cumulants,
\begin{equation}
\hat C_n^B = 2^n \hat C_n^p .
\end{equation}
This relation, however, holds only when correlations induced by Fermi statistics are negligible. While this condition is expected to be satisfied at high collision energies and temperatures~\cite{Kitazawa:2012at, Andronic:2017pug}, it may break down at low energies, where hadronic interactions dominate and quantum statistical effects become important, particularly near the nuclear liquid-gas phase transition~\cite{STAR:2022vlo}.

Factorial cumulants can be expressed as linear combinations of the ordinary susceptibilities~\cite{Bialas:2007ed, Bzdak:2016sxg, Friman:2025swg}. The first four factorial cumulants read:
\begin{align}
    \hat C^\alpha_1 &= \hat \chi_1^\alpha \textrm, \label{eq:fac_cum_def1}\\
    \hat C^\alpha_2 &= \hat \chi_2^\alpha - \hat\chi_1^\alpha \textrm,
\end{align}
where $\alpha = B, p, n$. Typically, factorial cumulants, $\hat C_n$, are defined as linear combinations of ordinary cumulants, $\hat \kappa_n$, which are related to the susceptibilities through $\kappa_n = VT^3\hat \chi_n$. Therefore, the definition of factorial cumulants differs by a factor $VT^3$ from the usual definition based on ordinary cumulants. Nevertheless, our definition does not lead to qualitative differences and the prefactors cancel out when ratios of cumulants are considered. We note that the correlation functions $\hat C^\alpha_k$ are defined for particles and not the net-particle numbers. However, in the vicinity of the liquid-gas phase transition, the contribution of antiparticles is well suppressed and we can stick to the relations above. Nevertheless, a direct relation between the factorial cumulants and susceptibilities for net-particle numbers can also be derived~\cite{Bzdak:2016sxg}. In the isospin-symmetric system, $\hat C_k^p = \hat C_k^n$.

The second-order baryon number factorial cumulant at $T=30~\mathrm{MeV}$ is shown in Fig.~\ref{fig:C2}. It exhibits a peak in the vicinity of the remnant of the liquid-gas phase transition and becomes negative at larger baryon chemical potential, reflecting the increasing role of repulsive interactions. The corresponding proton factorial cumulant, $\hat C_2^p$, shows a qualitatively similar behavior. Although $\hat C_2^B \simeq \hat C_2^p$, their difference displays a nontrivial structure.

To quantify this difference, we define
\begin{align}
\hat C_1^{pn} \equiv \hat C_1^B - \sum_{\alpha=p,n}\hat C_1^\alpha &= 0 ,\label{eq:frac_diff1} \\
\hat C_2^{pn} \equiv \hat C_2^B - \sum_{\alpha=p,n}\hat C_2^\alpha &= \hat \chi_2^{pn}. \label{eq:frac_diff2}
\end{align}
Thus, the difference between $\hat C_2^B$ and the factorial cumulants of individual species is directly given by the proton-neutron correlator $\hat \chi_2^{pn}$. A nonvanishing $\hat C_k^{pn}$ signals physics beyond an uncorrelated gas of nucleons, as expected near the liquid-gas phase transition.

The relations in Eqs.\eqref{eq:frac_diff1} and~\eqref{eq:frac_diff2} are exact and do not rely on additional assumptions; they hold for any temperature and baryon chemical potential. In general, a nonzero $\hat C_k^{pn}$ breaks the scaling relation $\hat C_k^B = 2^k \hat C_k^p$. This scaling could only be preserved if $\hat C_k^{pn} = (2^k - 2)\hat C_k^p$, which we do not find to be the case. Instead, $\hat C_k^{pn}$ is not proportional to $\hat C_k^p$, as illustrated in Fig.\ref{fig:C2}.

Factorial cumulants eliminate self-correlations present in ordinary susceptibilities, allowing genuine correlations to be isolated. We find that interaction-induced correlations, encoded in $\hat \chi_k^{pn} \neq 0$, modify the expected scaling between baryon and proton factorial cumulants. Although baryon and proton factorial cumulants exhibit qualitatively similar behavior near the liquid-gas transition, their difference remains nontrivial and displays comparable critical structures.

Finally, we note that ratios of factorial cumulants exhibit analogous behavior. While baryon and proton factorial cumulant ratios show qualitatively similar structures in the vicinity of the liquid-gas phase transition, their differences remain nontrivial and are governed by interaction-induced proton-neutron correlations. As in the case of the cumulants themselves, deviations from the simple scaling between baryon and proton factorial cumulant ratios directly reflect genuine correlations in the system. This further emphasizes that correlation effects persist not only at the level of individual cumulants, but also in their ratios, which are commonly employed in experimental analyses.

\section{Summary}
\label{sec:summary}

We studied net-baryon number density fluctuations in isospin-symmetric matter near the nuclear liquid-gas phase transition using the parity doublet model in the mean-field approximation. We analyzed susceptibilities of net-proton and net-neutron numbers, as well as their correlations. While these observables exhibit nonmonotonic behavior at the phase boundary, we find that proton-neutron correlations play a crucial role, leading to qualitative differences between net-proton and net-baryon fluctuations.

We further investigated factorial cumulants and showed that the qualitative structure of baryon-number factorial cumulants is largely preserved at the proton level. The difference between baryon- and proton-number factorial cumulants is governed by proton-neutron correlations, which display a similar nontrivial structure near the liquid-gas transition.

Our results highlight the importance of interaction-driven correlations and demonstrate that net-proton observables cannot, in general, be directly identified with net-baryon fluctuations near the liquid-gas critical endpoint. Although obtained within a mean-field framework, the emergence of proton-neutron correlations is a generic consequence of baryonic interactions, suggesting qualitative robustness of our conclusions.\\

\section*{Acknowledgment}
This work was partly supported by the program \textit{Excellence Initiative-Research University} of the University of Wroc\l{}aw, funded by the Ministry of Education and Science.

\end{document}